\documentclass{sigplanconf}

\usepackage[final]{listings}
\usepackage{color}
\usepackage{mathptmx}
\usepackage{graphicx}
\usepackage{color}
\usepackage{amsmath}
\usepackage{amsfonts}
\usepackage{amssymb}
\usepackage{threeparttable}
\usepackage{booktabs}

\begin{document}

\setlength{\pdfpageheight}{\paperheight}
\setlength{\pdfpagewidth}{\paperwidth}

\conferenceinfo{WPMVP '14}{February 15--19 2014, Orlando, FL, USA}
\copyrightyear{2014}
\copyrightdata{978-1-4503-2653-7/14/02}
\doi{2568058.2568068}
\newcommand{\comment}[1]{\color{red}!!! {#1} !!!\color{black}}

\title{Comparing the Performance of Different x86 SIMD Instruction Sets for
a Medical Imaging Application on Modern Multi- and Manycore Chips}

\authorinfo{Johannes Hofmann}
           {Chair for Computer Architecture\\University Erlangen--Nuremberg}
           {johannes.hofmann@fau.de}
\authorinfo{J. Treibig \and G. Hager \and G. Wellein}
           {Erlangen Regional Computing Center\\University Erlangen--Nuremberg}
           {jan.treibig@rrze.fau.de}

\maketitle

\begin{abstract}
    Single Instruction, Multiple Data (SIMD) vectorization is a major driver of
    performance in current architectures, and is mandatory for achieving good
    performance with codes that are limited by instruction throughput. We
    investigate the efficiency of different SIMD-vectorized implementations of
    the RabbitCT benchmark. RabbitCT performs 3D image reconstruction by back
    projection, a vital operation in computed tomography applications. The
    underlying algorithm is a challenge for vectorization because it consists,
    apart from a streaming part, also of a bilinear interpolation requiring
    scattered access to image data. We analyze the performance of SSE
    (128\,bit), AVX (256\,bit), AVX2 (256\,bit), and IMCI (512\,bit)
    implementations on recent Intel x86 systems. A special emphasis is put on
    the vector gather implementation on Intel Haswell and Knights Corner
    microarchitectures. Finally we discuss why GPU implementations perform much
    better for this specific algorithm.
\end{abstract}

\keywords
SIMD, Intel MIC, gather, computed tomography, back projection, performance

\section{Introduction}
\label{sec:intro}

Single Instruction, Multiple Data (SIMD) is a data-parallel execution model
enabling to do more work with the same number of instructions. Because of its
integration in existing microarchitectures it is a standard technology for
increasing performance in modern processors. While the idea
originated in the early seventies, the first implementation in a commodity processor
was Intel's MMX instruction set extension for the x86 Pentium processor line in
1996. With AltiVec/VMX (1998) for PowerPC and SSE (1999) for x86 more capable
128\,bit-wide SIMD instruction set extensions followed. For the software the
introduction of SIMD was a paradigm shift as in contrast to previous hardware
optimizations, which were designed to work transparently, the software must
make explicit use of the new instructions in order to leverage the full
potential of new processors. Vendors promised that optimizing compilers would
enable the efficient use of the new instructions, but it soon turned out that
in many cases data structures must be adapted and kernels must be rewritten
with the help of compiler intrinsics or even in assembly code. Intel
incrementally updated the SIMD capabilities by introducing new instruction set
extensions. SSE comprised 70 instructions, SSE2 (2001) added 144, SSE3 (2004)
29, and SSE4 (2006) 54 new instructions, for a total of 297 instructions
\cite{Inteldev}. In 2011 Intel doubled the SIMD width to 256\,bit with AVX.
Still of the total 349 instructions introduced with AVX only a subset of 92
instructions supported the 256\,bit registers and only few new instructions
were added. This was changed in 2013 with AVX2, which promoted most
instructions to 256\,bit and again added new ones. The new
instructions can be grouped into complex instructions for arithmetic
targeted at specific application classes, horizontal operations, more powerful
in-register reordering instructions, and operations promoted to vectorized
execution. While most of this can be seen as incremental improvements, AVX2
introduces instructions that change the way SIMD can be employed. This is the
\verb+gather+ instruction together with the ability to mask out SIMD lanes with
predicate registers. The hope is that this enables the use of SIMD in new
application classes with more complex data access patterns where SIMD could not
be used efficiently up to now. The first implementation of the AVX2 instruction
set is in the Intel Haswell microarchitecture. Haswell also features Fused
Multiply-Add (FMA) instructions in a separate instruction set extension (FMA3).

In 2006 Intel started developing an x86 many-core design (codename Larrabee),
initially targeted as an alternative to existing graphics processors. It
uses a 512\,bit SIMD instruction set called IMCI (Initial Many Core
Instructions). While in 2010 the usage as a GPU was terminated, the design was
refined and eventually re-targeted as an accelerator card. The first product
based on this design, the Intel Xeon Phi, was available in early 2013 with 60
cores and 8\,GB of main memory. IMCI already anticipated and implemented many
current developments in SIMD instruction sets and allowed to gain experience
with a working implementation. The instruction set comprises gather/scatter
instructions, predicate registers, and FMA.

This paper studies the efficiency of an important algorithm from medical
imaging when mapped to different Intel SIMD instruction set extensions. We look
at the Instruction Set Architecture (ISA) interface as well as its
implementation in a concrete microarchitecture. Although we also present
results for compiler-generated code, a detailed analysis of the conditions
under which a compiler would be able to achieve the same performance level as
hand-crafted assembly is out of scope for this paper. Still we discuss how well
an instruction set extension is suited for automatic generation of efficient
code.

We have chosen the RabbitCT benchmarking framework as test case
\cite{rohkohl2009}. RabbitCT implements volume reconstruction by back
projection, which is the performance limiting part of many Computed Tomography
(CT) applications. This operation is well suited for evaluating SIMD
implementations, because it consists of streaming as well as scattered data
access patterns and is therefore non-trivial to vectorize. Although a naive
performance model suggests that the runtime of the algorithm should be
determined by main memory bandwidth, its implementation with existing
instruction sets introduces so much overhead that instruction throughput
becomes the bottleneck. Therefore it can be seen as a prototype for
applications in which the data cannot be loaded block-wise, but instead
requires in-register shuffles. An interesting aspect of RabbitCT is that it is
also a popular target for GPU-based efforts and therefore provides a good case
to compare general purpose multi- and manycore designs with GPUs.

We picked an Intel Xeon IvyBridge-EP two-socket system for executing the SSE
and AVX code variants and an Intel Xeon Phi accelerator card for the IMCI
kernels. Unfortunately a multi-socket system implementing AVX2 and FMA3 is not
yet available. Therefore we choose a single socket Intel Xeon Haswell system to
execute the AVX2/FMA3 kernel.

This paper is structured as follows. Section \ref{sec:relatedwork} gives an
overview of previous work about the performance optimization of this algorithm.
The test systems used are described in Section
\ref{sec:testbed}. Section \ref{sec:rabbitct} introduces the RabbitCT benchmark
and motivates its use for this study. Next we cover the SIMD kernel
implementations in Section \ref{sec:implementation}. We briefly introduce all
optimizations applied and specifically cover the SIMD vectorization with an
emphasis on the gather instruction in IMCI and AVX2. The section closes with a
detailed instruction code analysis for the resulting kernels. The results are
presented in Section \ref{sec:results}. We separate single core from full
system results to be able to clearly analyze the SIMD speedup efficiency. We
also present microbenchmarking results showing the instruction latency for
different settings in the current gather implementations. In Section
\ref{sec:gpu_comparison} we compare our results with the fastest published GPU
result and try to explain why this specific application is well
suited for GPUs. We also present performance numbers for generated code in
Section \ref{sec:compiler}. To conclude we summarize our results and give an
outlook.

\section{Related Work}
\label{sec:relatedwork}

Due to its relevance in medical applications, reconstruction in computed
tomography is a well-studied problem. As vendors for CT devices are constantly
looking for ways to speed up the reconstruction time, many computer
architectures have been evaluated over time. Initial products in this field
used special purpose hardware based on FPGAs (Field Programmable Gate Arrays)
and DSPs (Digital Signal Processors) \cite{siemensfpga}. The Cell Broadband
Engine, which at the time of its release provided unrivaled memory bandwidth,
was also subject to experimentation \cite{cell1,cell2}. It is noteworthy that
CT reconstruction was among the first non-graphics applications that were run
on graphics processors \cite{ctwasfirst}. However, the use of varying data sets
and reconstruction parameters limited the comparability of all these
implementations. In an attempt to remedy this problem, the RabbitCT framework
\cite{rohkohl2009} provides a standardized, freely available CT scan data set
and a uniform benchmarking interface that evaluates both reconstruction
performance and accuracy. Current noteworthy entries in the RabbitCT ranking
include \textit{Thumper} by Zinsser and Keck \cite{thumper}, a Kepler-based
implementation which currently dominates all other implementations, and
\textit{fastrabbit} by Treibig et al. \cite{fastrabbit}, a highly optimized
CPU-based implementation. The presented work is partly based on the results of
\textit{fastrabbit}, improving it and providing an implementation for the Intel
IMCI and AVX2 instruction sets.

\section{Experimental Testbed}
\label{sec:testbed}
\begin{table*}[tb]
    \centering
    {\small
    \renewcommand{\arraystretch}{1.1}
	\begin{tabular}{lccc}
        \toprule
	    Microarchitecture           &Intel IvyBridge-EP&Intel Haswell    &Intel Knights Corner \\
	    Model                       &Xeon E5-2660 v2   &Xeon E3-1240 v3  &Xeon Phi 5110P\\
        \midrule
	    Base/Max. Turbo Clock Speed       &2.2\,GHz/3.0\,GHz              &3.4\,GHz/3.8\,GHz              &1.053\,GHz/--    \\
        Sockets/Cores/Threads per Node       &2/20/40           &1/4/8           &1/60/240           \\
	    SIMD Support                &SSE (128\,bit), AVX (256\,bit) &AVX2 (256\,bit), FMA3 (256\,bit)      &IMCI (512\,bit)\\
	    Vector Register File &16 Registers                &16 Registers               &32 Registers \\
        \midrule
	    Node L1/L2/L3 Cache              &20$\times$32\,kB/20$\times$256\,kB/2$\times$25\,MB    &4$\times$32\,kB/4$\times$256\,kB/8\,MB    &60$\times$32\,kB/60$\times$512\,kB/--  \\
	    Node Main Memory Configuration           &8 channels DDR3-1866   &2 channels DDR3-1600   &16 channels GDDR5 5GHz\\
        Node Peak Memory Bandwidth &119.4\,GB/s         &25.6\,GB/s                &320\,GB/s               \\
        Node Update Benchmark Bandwidth   &98\,GB/s (81\%)         &23\,GB/s (90\%)         &168\,GB/s (53\%)   \\
        \bottomrule
	\end{tabular}
    \caption{Test machine specifications.
    ``Node Update'' is a streaming multi-threaded benchmark that modifies an array in memory.}
    \label{tab:arch}
    }
\end{table*}
A standard two-socket server based on the IvyBridge-EP microarchitecture was
chosen for executing the SSE/AVX kernels. It employs two-way SMT and has ten
moderately clocked (2.2\,GHz base frequency) cores per socket. Sixteen vector
registers are available for use with SSE and AVX. Using floating-point arithmetic,
each core can execute
one multiply and one add instruction per cycle, leading to a peak performance
of eight Double
Precision (DP) or 16 Single Precision (SP) Flops per cycle. Memory bandwidth
is provided by means of a ccNUMA memory subsystem with four DDR3-1600 
memory channels per socket.

For the AVX2/FMA3 kernel a single-socket Intel Xeon server based on the Haswell
microarchitecture was used. This system has four cores (two-way SMT) and 8\,MB
of L3 cache.

The Intel Xeon Phi is located on a PCIe card and runs its own operating system.
It consists of a single chip providing 60 low-frequency cores with four-way
SMT. The cores are based on a modified version of P54C design used in the
original Pentium processor (1995). Each core is in-order and two-way
superscalar, featuring a scalar pipeline (V-pipe) and a vector pipeline
(U-pipe). The Vector Processing Unit (VPU) connected to the U-pipe features a
total of 32 512\,bit vector registers and is capable of FMA operations,
yielding a total of 16 DP (32 SP) Flops per cycle. The cores are connected via
a bidirectional ring bus. Eight GDDR5 memory controllers with two channels each
are connected to the ring bus. The Xeon Phi offers different settings for
usage, such as offload (common for GPUs) and native mode. We have used the
native execution model in which everything is executed on the accelerator card
and data transfers to and from the card are not taken into account. An overview
of the test machine specifications is given in Table \ref{tab:arch}.
\section{RabbitCT Benchmark}
\label{sec:rabbitct}

In CT an X-ray source and a flat-panel detector positioned on opposing ends of
a gantry move along a defined trajectory---mostly a circle---around the
volume that holds the object to be investigated. X-ray images are taken at
regular angular increments along the way. In general 3D image reconstruction
works by back projecting the information recorded in the individual X-ray
images (also called projection images) into a 3D volume, which is made up of
individual voxels (volume elements). In medical applications, this volume
almost always has an extent of $512^3$ voxels. To obtain the intensity value
for a particular voxel of the volume from one of the recorded projection images
we forward project a ray originating from the X-ray source through the center
of the voxel to the detector; the intensity value at the resulting detector
coordinates is then read from the recorded projection image and added to the
voxel. This process is performed for each voxel of the volume and all recorded
projection images, yielding the reconstructed 3D volume as the result.

Performance comparisons of different optimized back projection implementations
found in the literature can be difficult because of variations in data
acquisition and preprocessing, as well as different geometry conversions and
the use of proprietary data sets. The RabbitCT framework was designed as an
open platform that tries to remedy these problems. It features a benchmarking
interface, a prototype back projection implementation, and a filtered, high
resolution CT dataset of a rabbit; it also includes a reference volume that is
used to derive various image quality metrics. The preprocessed dataset
consists of 496 projection images that were acquired using a commercial C-arm
CT system. Each projection is 1248$\times$960 pixels wide and contains the
filtered and weighted intensity values as single-precision floating-point
numbers. In addition, each projection image $p_i$ comes with a separate
projection matrix $A_i \in \mathbb{R}^{3\times4}$, which is used to perform the
forward projection. The framework takes care of all required steps to set up
the benchmark, so the programmer can focus entirely on the actual back
projection implementation, which is provided as a module (shared library) to
the framework.

\subsection{Algorithm}

\lstset{
        breaklines=true,
        language=C,
        basicstyle=\small\ttfamily,
        numbers=left,
        numberstyle=\tiny,
        frame=t,
        columns=fullflexible,
        showstringspaces=false,
        escapechar=\%,
        escapebegin=\color{blue}\ttfamily\bfseries,
        keepspaces=true
}

\begin{lstlisting}[caption=Unoptimized reference back projection
implementation processing a single projection image.,
    label=src:fdk,
    float=htpb,
    captionpos=b,
    belowcaptionskip=4pt,
    ]
for (z=0; z<L; ++z) { // iterate over all voxels%\label{l:z}%
  for (y=0; y<L; ++y) {
    for (x=0; x<L; ++x) {%\label{l:x}%
      // PART 1
      // convert from VCS to WCS
      float wx = O + x * MM;%\label{l:wx}%
      float wy = O + y * MM;
      float wz = O + z * MM;%\label{l:wz}%
      // convert from WCS to ICS
      float u = wx*A[0] + wy*A[3] + wz*A[6] + A[9];%\label{l:u}%
      float v = wx*A[1] + wy*A[4] + wz*A[7] + A[10];
      float w = wx*A[2] + wy*A[5] + wz*A[8] + A[11];%\label{l:w}%
      // de-homogenize
      float ix = u / w;%\label{l:ix}%
      float iy = v / w;%\label{l:iy}%
      // integer indices to access projection image
      int iix = (int)ix;%\label{l:iix}%
      int iiy = (int)iy;%\label{l:iiy}%
      // calculate interpolation weights
      float scalex = ix - iix;%\label{l:scalex}%
      float scaley = iy - iiy;%\label{l:scaley}%
      // PART 2
      // load four values for bilinear interpolation
      float valbl = 0.0f; float valbr = 0.0f;%\label{l:zs}%
      float valtr = 0.0f; float valtl = 0.0f;%\label{l:ze}%
      if (iiy>=0 && iiy<width && iix>=0 && iix<height)%\label{l:vs}%
        valbl = I[iiy * width + iix];
      if (iiy>=0 && iiy<width && iix+1>=0
                              && iix+1<height)
        valbr = I[iiy * width + iix + 1];
      if (iiy+1>=0 && iiy+1<width && iix>=0
                                  && iix<height)
        valtl = I[(iiy + 1) * width + iix];
      if (iiy+1>=0 && iiy+1<width && iix+1>=0
                                  && iix+1<height)
        valtr = I[(iiy + 1)* width + iix + 1];%\label{l:ve}%
      // PART 3
      // perform bilinear interpolation
      float valb = (1-scalex)*valbl + scalex*valbr;%\label{l:bis}%
      float valt = (1-scalex)*valtl + scalex*valtr;
      float val = (1-scaley)*valb + scaley*valt;%\label{l:bie}%
      // add weighted results to voxel
      VOL[z*L*L+y*L+x] += val / (w * w);%\label{l:wt}%
    } // x-loop
  } // y-loop
} // z-loop
\end{lstlisting}

A slightly compressed version of the unoptimized reference implementation that
comes with RabbitCT is shown in Listing~\ref{src:fdk}. This code is called once
for every projection image. The three outer \texttt{for} loops (lines
\ref{l:z}--\ref{l:x}) are used to iterate over all voxels in the volume; note
that we refer to the innermost \texttt{x}-loop, which updates one ``line'' of
voxels in the volume, as the ``line update kernel.'' The loop variables \texttt{x},
\texttt{y}, and \texttt{z} are used to logically address all voxels in memory.
To perform the forward projection these logical coordinates used for addressing
must first be converted to the World Coordinate System (WCS), whose origin
coincides with the center of the voxel volume; this conversion happens in lines
\ref{l:wx}--\ref{l:wz}. The variables \texttt{O} and \texttt{MM} that are
required to perform this conversion are precalculated by the RabbitCT framework
and made available to the back projection implementation in a \texttt{struct}
pointer that is passed to the back projection function as a parameter.

After this the forward projection is performed using the corresponding
projection matrix in lines \ref{l:u}--\ref{l:w}. In order to transform the
affine mapping that implements the forward projection into a linear mapping
homogeneous coordinates are used. Thus the detector coordinates are obtained
in lines~\ref{l:ix} and~\ref{l:iy} by dehomogenization.

The next step is a bilinear interpolation, which requires converting the
previously obtained detector coordinates from floating-point to integer (lines
\ref{l:iix}--\ref{l:iiy}) to address the intensity values in the projection
image buffer \texttt{I}. The interpolation weights \texttt{scalex} and
\texttt{scaley} are calculated in lines \ref{l:scalex}--\ref{l:scaley}.

The four intensity values needed for the bilinear interpolation are fetched
from the buffer containing the intensity values in lines
\ref{l:vs}--\ref{l:ve}. The \texttt{if} statements ensure that the image
coordinates lie inside the projection image; in case the ray does not hit the
detector, i.e., if the coordinates lie outside the projection image, an
intensity value of zero is assumed (lines \ref{l:zs}--\ref{l:ze}). Note that
the projection image is linearized, which is why we need the projection image
width in the variable \texttt{width} (also made available by the framework via
the \texttt{struct} pointer passed to the function) to correctly address data
inside the buffer.

The actual bilinear interpolation is performed in lines
\ref{l:bis}--\ref{l:bie}. Before the result is written back into the volume, it
is weighted according to the inverse-square law (line \ref{l:wt}). The variable
\texttt{w} holds the homogeneous coordinate, which is an approximation of the
distance from the X-ray source to the voxel under consideration, and can be
used to perform the weighting.

For all further analysis we will structure the overall algorithm in three main
parts. Part 1 consists of the geometry computation involving the calculation of
the index in detector coordinates. Part~2 involves of the actual loading of the
intensity values from the projection image. Finally in Part 3 the bilinear
interpolation and the update of the voxel data is performed.

\subsection{Code Analysis}

One voxel sweep incurs a data transfer volume which consists of the loads from
the projection image and an update operation (\verb.VOL[i]+=s. for all
\verb.i.) to the voxel array. The latter causes 8\,bytes\ of traffic per voxel
and for the medically relevant problem size of $512^3$ voxels results in a data
volume of $1$\,GB per projection image, or $496$\,GB for all projections. The
main memory traffic caused by loading the intensity values from the projection
images is hard to quantify since it is not a simple stream; it is defined by a
``beam'' of locations slowly moving over the projection image as the voxel
update loop nest progresses. It exhibits some temporal locality, which allows
for a certain degree of caching, since neighboring voxels are projected on
proximate regions in the image, but there may also be multiple streams with
large strides if the beam sweeps across the image orthogonal to the rows. On
the computational side, the basic version of this algorithm performs 13
additions, 5 subtractions, 17 multiplications, and 3 divides.

\section{Implementation}
\label{sec:implementation}


Task parallel programming is based on OpenMP: The voxel volume is segmented
into voxels planes that can be processed independent from each other. All line
update kernels (SSE, AVX, AVX2/FMA3, IMCI) are written in assembly language and
use all of the optimizations found in the original \textit{fastrabbit}
implementation \cite{fastrabbit}.


The presented work improves on these achievements. The original algorithm
calculating the clipping mask\footnote{For some projection angles several
voxels are not projected onto the flat-panel detector. For these voxels a zero
intensity is assumed. Such voxels can be ``clipped'' off by providing proper
start and stop values for each \texttt{x}-loop.} had minor flaws, which have
been remedied. For a $512^3$ volume we can reduce the number of voxels that
have to be processed by almost 10\% when using the improved instead of the
original clipping mask. Moreover parameter handling and instruction scheduling
inside the kernel was improved leading to an overall speedup of a factor of
1.25 compared to the original \textit{fastrabbit} implementation.

\subsection{SIMD Vectorization}

Part 1 of the algorithm is straightforward to vectorize. A viable optimization
on all architectures is to replace the divide with a reciprocal instruction.
The reciprocal is pipelined with a throughput of one cycle and has a lower
latency compared to the standard divide. On KNC the reciprocal provides full
accuracy; on the CPU, however, it has a reduced accuracy (11\,bits of
mantissa). Nevertheless, the quality of the resulting reconstruction is
similar to that of GPU implementations. As a result of Part 1, the detector
indices have been computed and are located in SIMD registers.

Part~2 is the most difficult part of the algorithm, because the required
detector values are not contiguous in memory. Moreover, the contents of a SIMD
register cannot be used for addressing (i.e., as pointers or offsets) in SSE
and AVX. Therefore the SIMD register containing the indices must be stored back
to memory and loaded again into a general purpose register for addressing.
Since two adjacent values are always needed (cf. bilinear interpolation), the
image data can be loaded in pairs. The interpolation weights need to be
rearranged to match the order of the intensity values. While Part 2 looks
simple in C code, it requires a lot of instructions to implement it efficiently
using SSE and AVX. This part therefore does not scale very well with increased
register width.

In Part 3 the actual bilinear interpolation is performed. It requires some
additional reordering of the data in the SIMD registers, but is otherwise
straightforward to vectorize. The SSE and AVX implementations are very
similar, the latter requiring more reordering and adding more overhead for
loading the data due to its doubled SIMD width.

The main difference between SSE/AVX and AVX2/IMCI is the availability of FMA
and gather instructions in the latter. A significant part of the arithmetic
operations can be mapped to FMAs. The gather instruction simplifies the
implementation considerably, since it enables using the indices in the SIMD
register directly for addressing. However, some peculiarities of the IMCI ISA
diminish this benefit. The AVX2 implementation is very similar to IMCI but
provides a much simpler interface to the gather instruction, saving 50\%
instructions compared to the IMCI version. Instruction set-wise, the gather
instruction is a main benefit for this algorithm, so we introduce it in more
detail in the following section.

\subsubsection{Gather Instruction Interfaces}
\label{sec:gather}

\lstdefinelanguage
   [x]{Assembler}
   [x86masm]{Assembler}
   {morekeywords={vmovaps, xor, vprefetchenta, vprefetche2,
   vgatherdps, jknzd, jkzd, vsubps, vmulps, vaddps}}
\lstset{
        breaklines=true,
        language=[x]Assembler,
        basicstyle=\small\ttfamily,
        numbers=left,
        numberstyle=\tiny,
        frame=t,
        columns=fullflexible,
        showstringspaces=false
}
\begin{lstlisting}[caption={Gather instruction interface in IMCI.},
    label=src:4:imci_4,
    firstnumber=1,
    float=htpb,
    captionpos=b,
    belowcaptionskip=4pt,
    keepspaces=true
    ]
        kxnor       k2, k2
..L100: vgatherdps  zmm13{k2}, [rdi + zmm17 * 4] %\label{l:100}%
        jkzd        k2, ..L101%\label{l:zd}%
        vgatherdps  zmm13{k2}, [rdi + zmm17 * 4]%\label{l:dps2}%
        jknzd       k2, ..L100%\label{l:nzd}%
..L101:%\label{l:101}%
\end{lstlisting}

Listing \ref{src:4:imci_4} shows the instruction code for one gather construct
with IMCI. The \texttt{vgatherdps} instruction loads 16 single-precision
values (or fewer, depending on which bits of the mask register---\texttt{k2} in
our example---are set) into a register from the 16 addresses specified in its
second argument. Instead of fetching all data with one instruction,
\texttt{vgatherdps} works by getting the data cache line-wise per invocation;
this means that every time the gather instruction is used, it will fetch only
one cache line (CL), load all the values that it is supposed to gather from it,
store them in the destination vector register, and finally zero out the bits of
the components that have been filled in the vector mask register. As a
consequence, the number of gather instruction depends on the distribution of
the data: If all data resides in one CL then one gather instruction is
sufficient; in the worst case, each value is located in a different CL, which
will require sixteen gather instructions. The \texttt{jkzd} instruction in line
\ref{l:zd} checks the contents of the vector mask register that was updated in
the line before by the \texttt{vgatherdps} instruction. If the mask register is
zero, i.e., all zero bits indicating that all data has been gathered, control
flow continues at label \texttt{..L101} in line \ref{l:101}; if the mask
register is non-zero, i.e., if one or more bits are set, indicating that there
is still data to be fetched, no jump is performed and the \texttt{vgatherdps}
in line \ref{l:dps2} is scheduled next. After more data has been gathered and
\texttt{k2} has been updated, the vector mask register is tested again for zero
by the \texttt{jknzd} in line \ref{l:nzd}. If there is still data to be
fetched, control flow will jump back to label \texttt{..L100} in line
\ref{l:100}, starting the gather construct all over. If all data has been
fetched, the construct is left. This two-stage strategy was obtained by
examining the compiler output for non-contiguous data access and turns out
to be faster than a simple loop with only one gather instruction. The authors
have as yet no plausible explanation for this.

\begin{lstlisting}[caption={Gather instruction interface in AVX2.},
    label=src:4:AVX_4,
    firstnumber=1,
    float=htpb,
    captionpos=b,
    belowcaptionskip=4pt,
    ]
    vpcmpeqw        ymm7, ymm7, ymm7
    vgatherdps      ymm15, [rdi + ymm11 * 4], ymm7
\end{lstlisting}
Listing \ref{src:4:AVX_4} shows the respective code for AVX2. In contrast to
IMCI, no loop construct is required and a single instruction is sufficient.
This allows for a very compact and elegant code. Instead of using dedicated
vector mask registers, AVX2 uses regular vector registers (supplied as third
argument) as mask for the operation.

Note that for our application we always gather all vector components without
checking whether the intensity values are located inside the projection images
or not. This is achieved by setting the vector (mask) registers to all one
bits (\texttt{kxnor}, \texttt{vpcmpeqw}). We found that copying the projection
images into a zero-padded buffer and removing the conditionals (cf.
lines~\ref{l:vs}--\ref{l:ve} in Listing~\ref{src:fdk}) resulted in a better
performance than setting the mask registers, which causes too much instruction
overhead.

\subsection{Instruction Code Analysis}
\label{sec:instranal}

The main benefit of SIMD is that a given amount of work can be finished with
fewer instructions than with scalar execution. Therefore the number of
instructions and their composition in terms of instruction types (memory,
in-register, arithmetic, etc.) are interesting metrics for our case. These are
influenced by the register width and by the availability of instruction types,
such as FMA or gather instructions. Table \ref{tab:instranalysis} shows an
overview for the three parts of this algorithm for all five hand-written
assembly implementations. Note that for Haswell in addition to an AVX2 version
we also devised an AVX/FMA3 implementation that uses Haswell's FMA3 instruction
but is otherwise identical to the AVX implementation---i.e. it uses no gather
instructions.

Part 1 is trivial to vectorize and exhibits good scalability: In
Table~\ref{tab:instranalysis} this is reflected by the fact that hardly any
additional instructions are required when switching from SSE to AVX. Apart from
reducing the total number of arithmetic instructions, the use of FMA has the
additional benefit of reducing memory instruction in Part 1: Because a FMA
doesn't require an intermediate register for the result of the multiplication,
register pressure and---as a consequence---the amount of register spilling is
lower. In contrast to the first part, the second part of the algorithm scales
poorly instruction-wise. Due to the sequential loads, the number of memory
operations almost doubles when switching from SSE to AVX. For IMCI the gather
instruction simplifies the implementation, because no additional in-register
reordering is necessary. In contrast to the slim AVX2 gather interface, the
bold IMCI loop interface cancels out the advantage instruction-wise and ends up
with an instruction count similar to AVX. In Part 3, while in theory simple to
vectorize, all non-gather implementations suffer from in-register reordering
overhead.

Overall, the overhead of switching from SSE to AVX is 19 instructions. IMCI
roughly ends up with the same instruction count as the SSE code. While we would
anticipate a lower instruction count for IMCI considering FMA and gather, the
fact prefetching has to be implemented in software and the design of the gather
interface cause a lot of instruction overhead. The AVX2 gather interface does
not suffer from this shortcoming and has the lowest instruction count.

\begin{table}
    \centering
    {\small
    \begin{tabular}{llccccc}
        \toprule
        & Type              & SSE         & AVX         & AVX2        & AVX/FMA3    & IMCI \\
        \midrule
        Part 1&Memory       & 4           & 3           & 0           & 3           & 0 \\
        &Arith.             & 17          & 17          & 12          & 12          & 15 \\
        & All               & \textbf{21} & \textbf{20} & \textbf{12} & \textbf{15} & \textbf{15} \\
        \midrule
        Part 2&Memory       & 18          & 34          & 4           & 34          & 16 \\
        &Shuffle            & 6           & 10          & 0           & 10          & 0 \\
        &Arith.             & 2           & 2           & 12          & 2           & 24 \\
        & All               & \textbf{26} & \textbf{46} & \textbf{16} & \textbf{46} & \textbf{40} \\
        \midrule
        Part 3 &Memory      & 2           & 2           & 2           & 2           & 2  \\
        &Arith.             & 20          & 20          & 15          & 16          & 12 \\
        & All               & \textbf{22} & \textbf{22} & \textbf{17} & \textbf{18} & \textbf{14} \\
        \midrule
        Other & All         & 4 & 4 & 4 & 3 & 8\\
        \midrule
        Total &             & \textbf{73} & \textbf{92} & \textbf{49} & \textbf{82} & \textbf{77} \\
        \bottomrule
    \end{tabular}
    \caption{Instruction count and composition. The instruction types are categorized into three classes: \emph{memory} for all instructions with a memory reference, \emph{shuffle} for register manipulation instructions, and \emph{arithmetic} for computational instructions. All remaining instructions such as, e.g., loop instructions are contained in \emph{other}.}
    \label{tab:instranalysis}
}
\end{table}

In Table \ref{tab:scalar-vs-simd} we compare the number of instructions
required by different SIMD instruction sets compared to baseline scalar code.
The baseline code was generated using the Intel 13.1 compiler together with the
\texttt{-O3 -no-vec -x} flag for each of the instruction sets. This means
that for AVX and AVX2 the scalar version makes use of the new VEX (vector
extensions) prefix introduced by AVX, allowing instructions to use more than
two operands; the AVX2 and AVX/FMA3 versions use FMA instructions. We omit
IMCI in this comparison, because there exist no dedicated scalar instructions
in this instruction set. The ``instruction count efficiency'' metric is the
ratio of instruction counts between the scalar and the vectorized code
versions---e.g. an efficiency of 100\% means that both versions require the
same amount of instructions, whereas an efficiency of 50\% means that the
vectorized code requires twice as many instructions as the scalar code.
Additionally we also list the ``SIMD runtime efficiency,'' which we define as
the achieved runtime speedup divided by the number of SIMD lanes. Interestingly,
the SIMD efficiency numbers for SSE and AVX  are almost identical to the
respective instruction count efficiencies. 
For both the AVX2 and AVX/FMA3 implementations the baseline scalar
code was generated using the \texttt{-xCORE-AVX2}, which is why the AVX/FMA3
code shows a lower runtime efficiency than AVX. Interestingly we find that
while instruction count efficiency is higher using AVX2, AVX/FMA3 provides
better runtime efficiency, i.e. performance. The reason for this is the high
latency of the gather instruction that is used in the AVX2 variant (cf.
Section~\ref{sec:gatherperf}).

\begin{table}
    \centering
    {\small
    \begin{tabular}{lcccc}
        \toprule
        &SSE &AVX & AVX2 & AVX/FMA3 \\
        \cmidrule(r){2-5}
        Voxels per Vectorized Loop & 4 & 8 & 8 & 8 \\
        Instr. per Loop (SIMD)   &73    &92   &49   & 82   \\
        Instr. per Voxel (Scalar)&57    &46   &41   & 46   \\
        Instr. Count Efficiency  &78\%  &50\% &84\% & 56\% \\
        SIMD Runtime Efficiency  &82\%  &51\% &33\% & 42\% \\
        \bottomrule
    \end{tabular}
    \caption{Overview of static instruction code analysis.  Runtime Efficiency
        results pertain to one core using SMT.}
        \label{tab:scalar-vs-simd}
}
\end{table}

\section{Results}
\label{sec:results}

Section \ref{sec:instranal} has already described the SIMD efficiency of
various implementations in terms of instruction overhead. Here we present
results illustrating how well the instruction set was implemented in the
microarchitectures and how efficient the implementation is in terms of speedup.
To do so we distinguish two cases: Single core (to pinpoint the SIMD influence
alone) and full-system scaling. It was shown previously \cite{fastrabbit} that
the performance of the code is limited by instruction execution, and that data
transfers through the cache hierarchy do not play a significant role on modern
multi-core architectures. Thread pinning was done with
likwid-pin \cite{likwid-psti}.

\subsection{Analysis of SIMD and SMT Speedup}

\begin{figure}[tb]
\centering
\includegraphics[clip=true,width=0.78\linewidth]{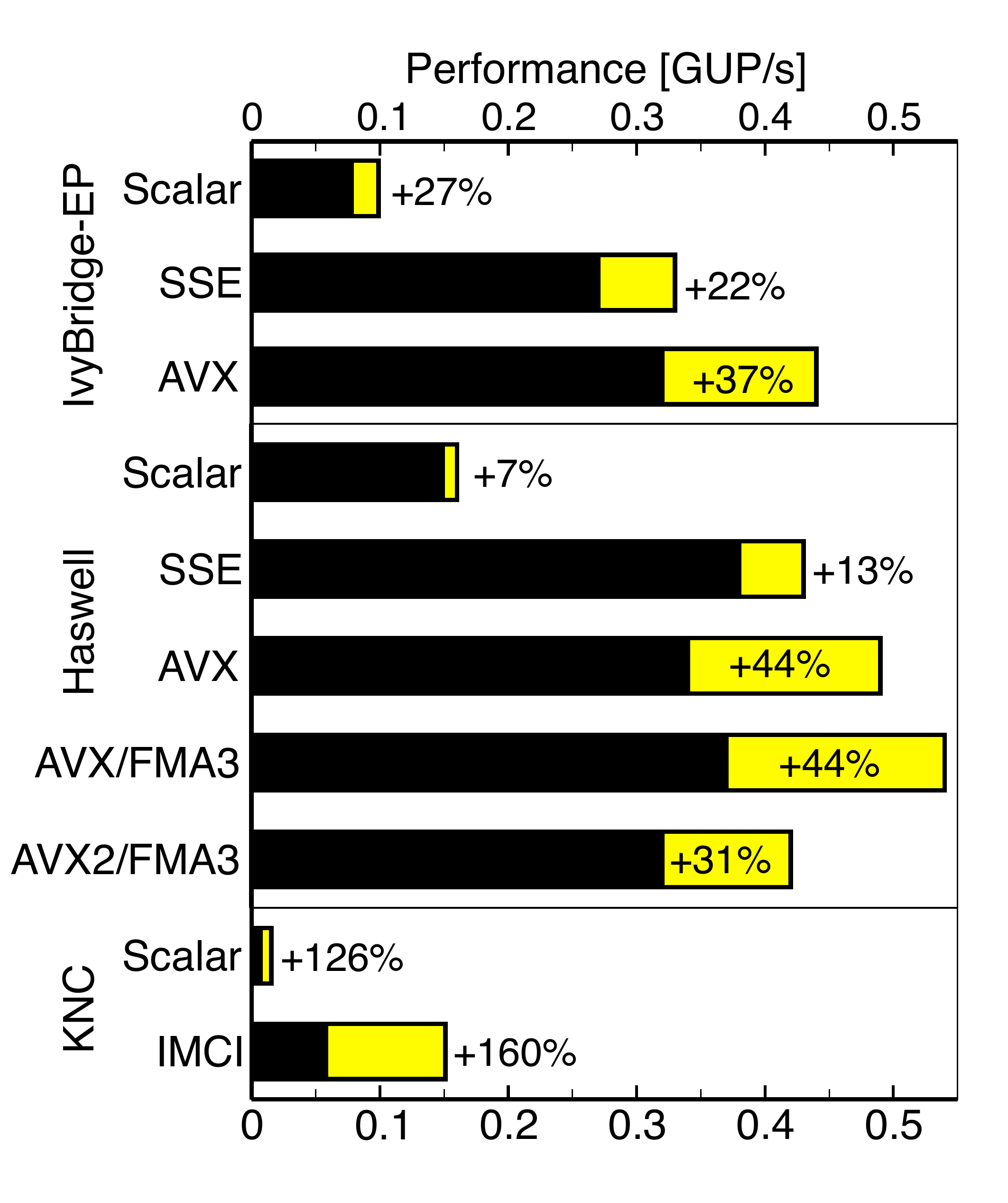}
\caption{Single core performance of different SIMD implementations on
IvyBridge-EP, Haswell and Knights Corner (KNC) in billions of
voxel updates per second (GUP/s). The black bars are
sequential results. The yellow bar is the speedup using all available SMT threads. The scalar version on IvyBridge-EP based on AVX and on Haswell on AVX2.}
\label{fig:simd}
\end{figure}

Figure \ref{fig:simd} shows the results for single core performance. Again,
scalar code was generated using the Intel 13.1 compiler. It is instructive to
analyze the benefit of Simultaneous Multi-Threading (SMT) together with SIMD,
since a considerable speedup with SMT points to inefficient pipeline
utilization. On IvyBridge-EP, SSE yields a 3.3$\times$ speedup with SMT (out of
a possible 4$\times$). As expected, the AVX kernel is significantly less
efficient, with 4.1$\times$ out of 8$\times$. It is known that the AVX kernel
suffers from critical path dependencies \cite{fastrabbit}, which is confirmed
by the large benefit gained with SMT. In a sense, inefficient SIMD code can be
partly compensated by multi-threading in this particular case. On Haswell the
SSE kernel performs surprisingly well, even without SMT, but the best possible
variant is AVX/FMA3 with SMT. The AVX2 kernel is the slowest of all variants
despite its instruction count advantage: The bilinear interpolation requires
four intensity values for each voxel, which results in four gather
instructions. These incur a latency of about 42~clock cycles (cf. Sec.
\ref{sec:gatherperf}). In contrast the AVX and AVX/FMA3 implementations use 16
pairwise loads to gather the intensity values; although these version require
about 10 additional in-register shuffling instructions overall this approach
yields a much better performance. On KNC the SIMD speedup is 10$\times$ out of
a possible 16$\times$. This result emerges from a combination of slow scalar
code (see above) with an inefficient and dominating gather instruction in the
SIMD case. We will analyze the latter in more depth in Section
\ref{sec:gatherperf}.

\subsection{Full Device Results}

\begin{figure}[tb]\centering
\includegraphics[angle=-90,clip=true,width=0.7\linewidth]{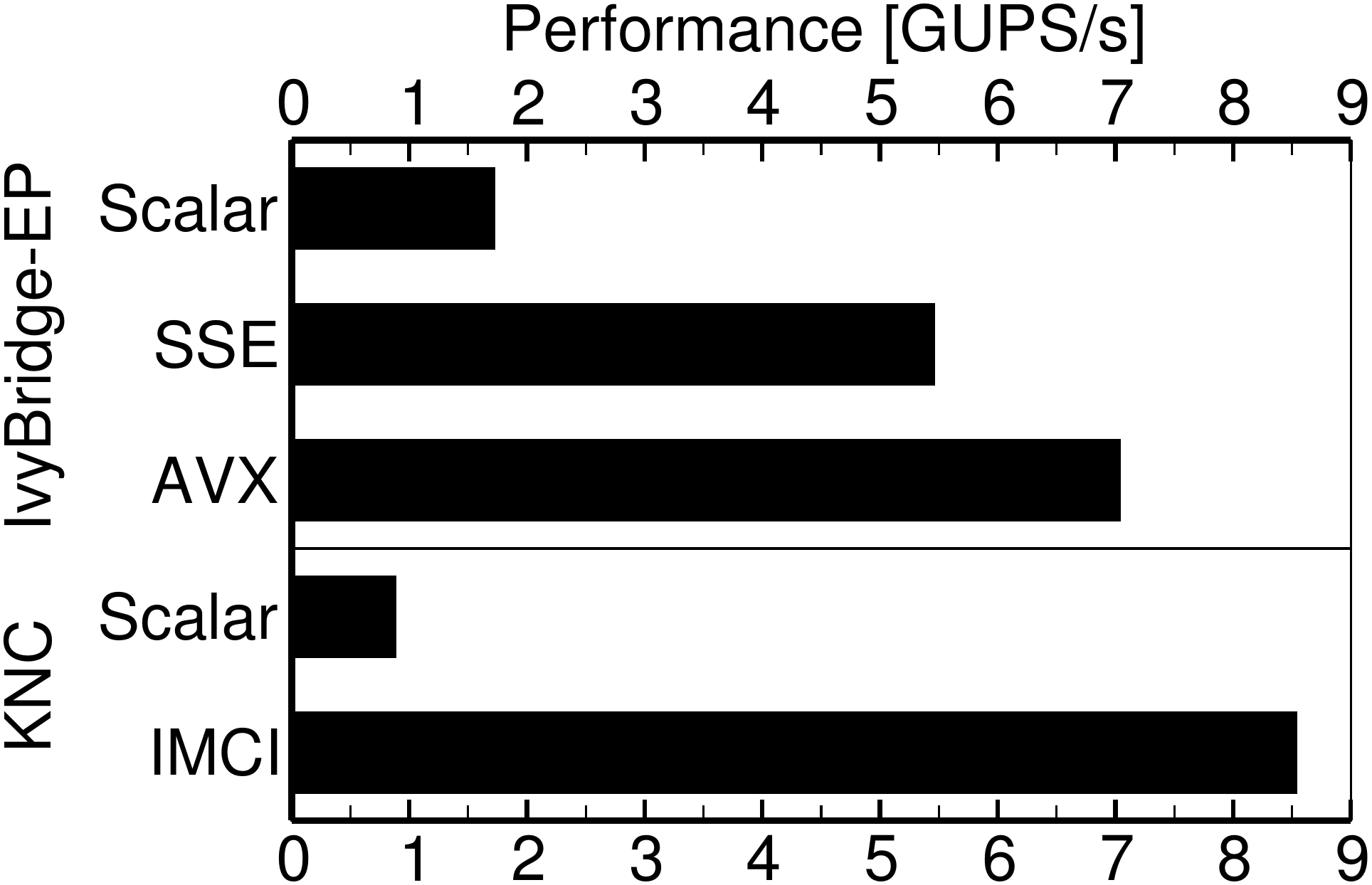}
\caption{Full system performance on IvyBridge-EP and KNC}
\label{fig:device}
\end{figure}

Figure \ref{fig:device} shows the performance of scalar and vectorized
implementations on IvyBridge-EP and KNC. Results for Haswell are omitted
because at the time of writing no two-socket system was available. The full
system results were measured with Turbo mode enabled and all available SMT
threads. A naive peak performance comparison gives reason to expect a 3$\times$
advantage for KNC. For our algorithm, one KNC device is only 1.2$\times$
faster than one IvyBridge-EP node. This is a disappointing result, considering
that IvyBridge-EP is also available with 12 instead of 10 cores and with a
faster clock speed. Both architectures suffer from the impact of the
inefficient Part 2 execution, but this problem is more severe on KNC, as will
be demonstrated in Section \ref{sec:perfmodel}.

Scalability across the cores is very similar on both systems, with a parallel
efficiency of 93\%.\footnote{To obtain results unbiased by Turbo mode artifacts
we disabled Turbo mode on IvyBridge during the scalability measurements.} This
corroborates the expectation that code execution is highly core-bound.

%

\subsection{Detailed Performance Analysis of Vector Gather Implementations}
\label{sec:gatherperf}

The vector gather instruction in the KNC microarchitecture has attracted major
attention as it is the first gather implementation in an x86-based design. In
this section we conduct a more detailed analysis of the gather performance on
KNC, and compare with the situation on Haswell (with AVX2).

As introduced in Section \ref{sec:gather} the KNC ISA requires the gather
instructions to be invoked multiple times in order to load all required data
elements. The exact number of executions depends on the total number of cache
lines the data is loaded from. Using the likwid-bench tool \cite{likwid-psti}
we implemented a microbenchmark which allows to measure the average instruction
latency depending on how many values are loaded per call (cache line). All
hardware threads of a single core were used to measure the latencies to gather
16 elements from L1 cache and L2 cache, depending on how the data is
distributed. The results are shown in Table~\ref{tab:2:gather_latency}.
\begin{table*}[tb]
  \begin{center}
{\small
    \begin{tabular}{ c r r r r r r }
    \toprule
    Microarchitecture & \multicolumn{4}{c}{Knights Corner} & \multicolumn{2}{c}{Haswell} \\
    \cmidrule(r){2-5}               \cmidrule(r){6-7}
     Distribution &   \multicolumn{2}{c}{L1 Cache} &  \multicolumn{2}{c}{L2 Cache}  & \multicolumn{1}{c}{L1 Cache} &  \multicolumn{1}{c}{L2 Cache} \\
    \cmidrule(r){2-3}               \cmidrule(r){4-5}          \cmidrule(r){6-6} \cmidrule(r){7-7}
                              & Instruction & Loop & Instruction & Loop  & Instruction & Instruction \\
    \midrule
        16 per CL               & 9.0   & 9.0   & 13.6     & 13.6      \\
        8 per CL               & 4.2   & 8.4   & 9.4     & 18.8        & 10.0 & 10.0 \\
        4 per CL               & 3.7   & 14.8   & 9.1     & 36.4       & 11.0 & 11.2 \\
        2 per CL               & 2.9   & 23.2   & 8.6     & 68.8       & 10.0 & 12.0 \\
        1 per CL               & 2.3   & 36.8   & 8.1     & 129.6      & 11.2 & 11.2 \\
    \bottomrule
    \end{tabular}
}
  \end{center}
  \caption{Latencies in clock cycles encountered when gathering data from
    different levels of the memory hierarchy for different distributions of the
    data to be gathered. ``Instruction'' and ``Loop'' refer to the average latency
    of a single \texttt{vgatherdps} instruction and the total time required to
    gather all 16 (IMCI) or 8 (AVX2) elements, respectively.}
  \label{tab:2:gather_latency}
\end{table*}

We find that the latency for a single \texttt{vgatherdps} instruction on KNC
varies depending on how many elements it has to fetch from a cache line. The
reason for this effect is that the \texttt{vgatherdps} instruction itself is
implemented as another loop, not visible on the ISA level: the more elements it
has to fetch from a single CL, the higher the latency. Nevertheless we find
that it is beneficial for the data to reside in as few CLs as possible:
Although the latency for a single \texttt{vgatherdps} instruction increases
with the number of elements per cache line, the impact of calling the
\texttt{vgatherdps} instruction multiple times is generally more severe.
On the Haswell microarchitecture, the gather latency is largely independent of
the number of cache lines touched when data is in the L1 or the L2 cache. From
the numbers it is evident that it may be faster in some cases to ignore the
hardware-based gather. The good performance of the AVX/FMA3 algorithm
implementation on Haswell compared to AVX2 (see Fig.~\ref{fig:simd}) is a
direct consequence of this.

On KNC, which has no L1 hardware prefetcher, there is a noticeable impact when
dropping from L1 to L2 cache. Due to the streaming access pattern in the
microbenchmark we find that on Haswell there is almost no difference in latency
depending on whether data is gathered from L1 or L2 cache due to hardware
prefetching. Overall, the performance impact of the gather operation is much
more severe on KNC than on Haswell. In Section~\ref{sec:perfmodel} we will show
that the gather instructions do indeed account for the dominant part of the
kernel runtime on KNC, leading to the observed mediocre KNC performance.

\subsection{Performance Analysis for Knights Corner}
\label{sec:perfmodel}


We start by estimating how many cycles are needed to execute
one loop iteration of the kernel. Neglecting the variable influence of the gather constructs
in Part 2 for the time being, we created a variant of the full kernel without
gather instructions.

Based on a static instruction code analysis taking into account instruction
pairing (superscalar execution) we estimate 34 cycles for the gather-less
kernel. This was verified by measurement which resulted in about 37.5 cycles.
For our model, we use the measured value of 37.5~clock cycles because it
contains non-negligible overhead such as backing up caller-save registers when
calling the line update kernel that was not accounted for in the analytical
prediction.

By instrumentation of the gather loops it was determined that for one loop
iteration the gather instruction was executed 16 times on average.
Distributing that number over the four gather loop constructs (one for each of
the four values required for the bilinear interpolation) we arrive at 4 gather
instructions per gather loop---indicating that the data is, on average,
distributed across four CLs. From this we can infer the runtime contribution
based on our previous findings (cf. Table~\ref{tab:2:gather_latency}). The
latency of each gather instruction in the situation where the data is
distributed across four CLs is 3.7~clock cycles. With a total of 16~gather
instructions per iteration, the contribution is 59.2~clock cycles. Together
with the remaining part of one kernel loop iteration (37.5~clock cycles), the
total execution time is approximately 97~clock cycles.

Up until now we assumed that all data is already located in the L1 cache.
Using likwid-perfctr \cite{likwid-psti} we 
found that 88.5\% of the projection data can be serviced from the
local L1 cache and the remaining 11.5\% can be serviced from the local L2
cache. Since each gather transfers a full CL, this amounts to
approximately $16\,\textrm{CLs} \cdot 64\,\textrm{byte/CL} \cdot 11.5\% \approx
118\,\textrm{byte}$. We estimate the \textit{effective} L2 bandwidth in
conjunction with the gather instruction to be the following: The latency of a
single gather instruction (with data distributed across four CLs) was
previously measured to be 3.7\,clock cycles with data in L1 cache, respectively
9.1\,clock cycles with data in the L2 cache (cf.
Table~\ref{tab:2:gather_latency}). Assuming the difference of 5.4\,clock
cycles to be the exclusive L2 cache contribution, we arrive at an effective
bandwidth of $64\,\textrm{byte} / 5.4\,\textrm{cycle} =
11.85\,\textrm{byte/cycle}$. The additional cost is thus $118\,\textrm{byte} /
11.85\,\textrm{byte/cycle}\approx10\,\textrm{cycles}$, resulting in a total
runtime of 107\,clock cycles.

Several unsuccessful attempts to improve the L1 hit rate of the gather
instructions were made. We found that the gather hint instruction,
\texttt{vgatherpf0hintdps}, is implemented as a dummy operation---it has no
effect whatsoever apart from instruction overhead. Another prefetching
instruction, \texttt{vgatherpf0dps}, appeared to be implemented exactly the
same as the actual gather instruction, \texttt{vgatherdps}: Instead of
returning control back to the hardware context after the instruction is
executed, we found that control was relinquished only \textit{after} the data
has been fetched into the L1 cache, rendering the instruction useless. Finally,
scalar prefetching using the \texttt{vprefetch0} instruction was evaluated.
Unfortunately the instruction overhead of moving the offsets from vector
registers onto the stack to get them into general purpose registers for scalar
prefetching far outweighed the benefit of improving the L1 hit rate (even when
prefetching only the CLs containing every second, forth, or even eight value).

To summarize, out of the 107~clock cycles 69 can be attributed to gathering the
required data, clearly indicating that the gather implementation is the
factor limiting SIMD scalability. If pairwise loads and an adequate latency
hiding mechanism were available, this could be reduced to a mere 32
cycles (2$\times$16 pairwise loads for the 16 voxels that are processed
in one loop iteration).

\section{Comparison to GPU Implementations}
\label{sec:gpu_comparison}

In order to integrate our findings with today's state-of-the-art in CT image
reconstruction, a comparison with the fastest currently available GPU
implementation called \textit{Thumper} \cite{thumper} shows that the GeForce
GTX 680 is almost 8$\times$ faster than KNC, although Intel had originally
intended the KNC to compete with GPU accelerators from other vendors. This
discrepancy can not be explained by simply examining the platforms'
specifications such as peak Flop/s and memory bandwidth.

Two of the main causes contributing to the GPU's superior performance in this
particular application are:
\begin{enumerate}
    \item Most computations involved in the reconstruction kernel, such as the
projection of voxels onto the detector panel or the bilinear interpolation, are
typical for graphics applications (which GPUs are designed for). While the
matrix-vector multiplication is performed efficiently on both the GPU and the
KNC, the bilinear interpolation is much faster on the GPU: GPUs have additional
hardware (texture units) that can perform multiple bilinear interpolations in
each clock cycle for data in the texture cache. To emphasize the
implications, consider that out of the total of 107~clock cycles for one loop
iteration of the kernel, 94~clock cycles, i.e., almost 90\%, are spent on the
bilinear interpolation, which can be performed with a single instruction on a
GPU.

    \item Given a sufficient amount of work, Nvidia's CUDA programming model
does a better job at hiding latencies. As seen before, even in the ideal case,
where all data can be serviced from the L1 cache, on average each of the gather
instructions has a latency of 3.7~clock cycles. Although the KNC can hide the
latencies of most instructions when using all four hardware contexts of a core,
4-way SMT is not sufficient to hide latencies caused by loading non-contiguous
data, and is still plagued by excess traffic due to the cache line concept. 
The massive threading on Nvidia's multiprocessors 
ensures that there is always a sufficient number of warps to choose from
when a particular warp stalls.
This approach can hide much longer latencies than the 4-way SMT in-order
approach of KNC.
\end{enumerate}

\section{Performance Comparison to Generated Code}
\label{sec:compiler}

\begin{figure}[tb]
\centering
\includegraphics[width=\linewidth]{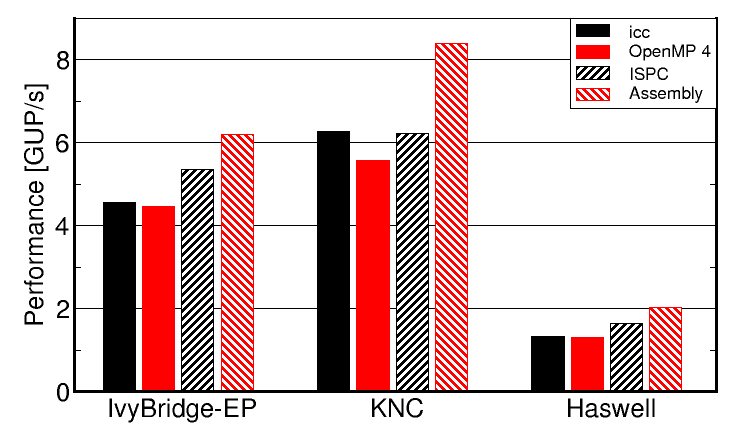}
\caption{Performance comparison of different programming models on full systems. }
\label{fig:results_gups}
\end{figure}


Figure \ref{fig:results_gups} compares our best implementations on the three
platforms with compiler-generated code. We consider three different variants: C
code, C code with the \verb.#pragma simd. directive from the latest OpenMP~4
standard \cite{omp4}, and ISPC \cite{ispc}. The Intel C compiler version 13.1.3
with optimization flags \verb+-O3 -xHost+ was used on all systems.
%
%
It is noteworthy that the performance using \texttt{\#pragma simd} is worse
than without the directive. ISPC (Intel SPMD Program Compiler, Version 1.5.0)
is an Open Source implementation of the SPMD programming model. It provides
the same performance as the Intel C compiler on KNC but on both CPUs it offers
performance superior to that of the commercial Intel compiler.

The hand-written assembly kernels outperform compiler-gen\-er\-ated code on
every architecture. On IvyBridge the AVX version is 10\% faster than the
ISPC-generated code. The effect is more pronounced on Haswell: Because the
ISPC-generated code makes use of the vector gather instruction, the hand-written
assembly using pairwise loads is 22\% faster. On KNC the Intel compiler
provided the fastest auto-vectorized code; the hand-written IMCI version is
34\% faster.



%
\section{Conclusion}
\label{sec:conclusion}

We have implemented the RabbitCT benchmark algorithm using different SIMD
instruction set extensions and have benchmarked the resulting kernels on three
recent Intel x86 architectures. The SIMD instruction sets exhibit different
register widths ranging from 128 to 512\,bits. Moreover, AVX2/FMA3 and IMCI
provide instructions for vector gather and FMA. Using an instruction code
analysis we have shown that it is not efficient to employ wider SIMD widths for
this algorithm due to its partially scattered data access pattern. We also show
that FMA and gather have a significant impact on the implementation, not only
with regard to instruction count but also in terms of simplicity. By far the
most compact and straightforward variant is the AVX2/FMA3 kernel. From the ISA
point of view we think this is instruction code which is well suited to be
automatically generated. We have then benchmarked the kernels to test the
hardware implementations. The advantages at the ISA level of the
gather-enabled SIMD instruction sets are currently thwarted by inefficient
hardware implementations. On KNC and Haswell the current gather throughput is
the dominating performance bottleneck for these kernels. An in-detail
microbenchmarking analysis on these two systems revealed that the
implementations suffer from significant overhead. Another issue is the fact
that there is no functional latency hiding for the gather operation on KNC.

Still the new instruction sets make it easier for a compiler to generate
competitive code. Therefore we think that the problem is solved on the ISA
side and the vendors have to provide improved implementations to further
increase the benefit of using these instructions. The advantage of GPUs for
this algorithm can be explained by the bilinear interpolation being implemented
completely in hardware. The second advantage is the more robust and easier to
use latency hiding strategies on GPUs compared to the available multi- and
many-core architectures.

We believe that RabbitCT is a very good benchmarking case to test the
efficiency of available instruction sets, code generators, and
microarchitectures. Future work will cover the port of our kernels on further
SIMD instruction sets such as IBM VSX.

\acks

We thank IBM Research for giving Jan Treibig the opportunity for a scientific
visit at the T.J.Watson Research Center, which was the starting point for this
work. Special acknowledgments go to Jose Moreira for fruitful discussions.

\bibliographystyle{abbrvnat}
\bibliography{rrze}

\end{document}